\documentclass[amstex,prl,twocolumn,aps,psfig,showpacs]{revtex4}
\usepackage{amsmath}
\usepackage{amssymb}
\usepackage{graphicx}

\begin{document}

\title{Fulde-Ferrel-Larkin-Ovchinnikov Inhomogeneous Superconducting State
and Phase Transitions Induced by Spin Accumulation in a Ferromagnet-$%
d_{x^{2}-y^{2}}$-Wave Superconductor-Ferromagnet Tunnel Junction}
\author{Biao Jin, Gang Su$^{\ast}$ and Qing-Rong Zheng }
\affiliation{College of Physical Sciences, Graduate University of Chinese Academy of
Sciences, P.O. Box 4588, Beijing 100049, China}

\begin{abstract}
Fulde-Ferrel-Larkin-Ovchinnikov (FFLO) inhomogeneous superconducting (SC)
state, first- and second-order phase transitions, and quantum criticality
induced by spin accumulation in a ferromagnet-$d_{x^{2}-y^{2}}$-wave
superconductor-ferromagnet tunnel junction are theoretically predicted. A
complex phase diagram in the temperature-bias voltage plane is determined.
It is found that the phase transitions from the homogeneous BCS state to the
inhomogeneous FFLO state, and from the FFLO state with the momentum $\mathbf{%
q}$'s azimuthal angle $\theta _{\mathbf{q}}=0$ to that with $\theta _{%
\mathbf{q}}=\pi /4$, are of the first-order; while the transitions from all
SC states to the normal state at critical voltages are of the second-order.
A Lifshitz point, a bicritical point and a quantum critical point are
identified.
\end{abstract}

\pacs{74.50.+r, 75.47.De, 73.40.Rw}
\maketitle

\textit{Introduction.} ---Forty years ago Fulde and Ferrel\cite{FF}, Larkin
and Ovchinnikov\cite{LO} (FFLO) pointed out independently that owing to the
presence of the spin-exchange field, the superconducting (SC) order
parameter in a ferromagnetic superconductor can be spatially modulated in
real space, leading to that the SC state inside these materials is
inhomogeneous. Such a spatially modulated SC state has long been detected in
a number of systems under various circumstances (e.g. \cite{casal,kontos}).
Nonetheless, a clear and direct confirmation of the existence of the FFLO
state has not yet been reported sofar, because a successful observation of
the FFLO state requires very clean type II superconductor with the
characteristic parameter $\kappa \gg 1$, the condition is not easily
satisfied in conventional superconductors. Recently, the possibility of the
FFLO states in ultracold atomic Fermi gases has been actively discussed (see
e.g. Refs. \cite{atomic}), and a new phase observed in CeCoIn$_{5}$ is also
found to be the FFLO state with a spatially modulated superconducting order
parameter\cite{Kakuyanagi}. On the other hand, for a magnetic sandwiched
heterostructure with a normal metal or superconductor as a spacer, when the
spin-polarized electrons driven by an external bias voltage from a
ferromagnetic film enter into the spacer faster than the spin-polarization
can diffuse away from the interface, there must be some nonequilibrium spin
densities built up near the interfaces of the heterostructures that depend
on the relative orientations of magnetic moments, leading to the occurrence
of spin accumulation. The spin accumulation is an important effect in the
emerging field of spintronics (see, e.g. Refs. \cite%
{meser,gijs,prinz,mood,wolf,sarma,book,zutic} for review), which plays a key
role in the spin Hall effect\cite{spin-hall} as well as in various
spintronic devices such as spin transistors, spin valves, spin diodes, spin
field-effect transistor, and so on. Quite recently, the spin accumulation
has been experimentally observed in a n-type GaAs by means of Kerr rotation%
\cite{kato}. Another interesting observation is that the spin accumulation
could result in the suppression of superconductivity, that happens in
ferromagnet/superconductor (F/S) heterostructures (e.g. Refs. \cite%
{takahashi,zheng,tserko,yoshida,chen,jin,johansson}). The cause is that the
injected spin-polarized electrons into the central superconductor give rise
to a spin imbalance, leading to a spin density accumulated near the
interfaces. Such a spin density is equivalent to a small magnetic field that
plays a role as a pair-breaking field, thereby enabling the
superconductivity to be suppressed. As the spin accumulation in the central
superconductor could generate an equivalent magnetic field, one has very
reasons to expect that the FFLO spatially modulated SC state should appear
in such a system.

There has been a recent study\cite{jin1} showing that in a certain range of
bias voltage the FFLO inhomogeneous SC state can indeed occur in the central
superconductor of a ferromagnet/s-wave superconductor double tunnel junction
in case of antiparallel alignments of the magnetizations of both
ferromagnets. It has been found that the phase transition from the
homogeneous s-wave BCS state to the inhomogeneous FFLO state is of
first-order, that gives rise to the oscillating behaviors of the tunnel
magnetoresistance (TMR) and conductance. These oscillating features can be
used as an alternative way to examine the existence of the FFLO state. In
this paper, we shall investigate the FFLO state induced by the spin
accumulation in a central $d_{x^{2}-y^{2}}$-wave superconductor of the
ferromagnet/d-wave superconductor double tunnel junctions. As the SC gap in
a d-wave superconductor is anisotropic, it would induce more exotic
behaviors than in a s-wave superconductor. It is shown that the phase
transitions from the homogeneous $d_{x^{2}-y^{2}}$-wave BCS state to the
inhomogeneous FFLO state, and from the FFLO state with the azimuthal angle
of the momentum $\theta _{\mathbf{q}}=0$ to that with $\theta _{\mathbf{q}%
}=\pi /4$, are of the first-order; while the transitions from all SC states
to the normal state at critical voltages are of the second-order. A complex
phase diagram in the temperature-bias voltage plane is determined, and a
Lifshitz point, a bicritical point and a quantum critical point are
identified.

\textit{Model and formalism.} ---We start with a symmetric ferromagnet/$%
d_{x^{2}-y^{2}}$-wave superconductor/ferromagnet (F/d-SC/F) double tunnel
junction with the left and right ferromagnetic (FM) electrodes applied by
bias voltages $V/2$ and $-V/2$, respectively. The two identical FM
electrodes with antiparallel alignment of both magnetizations are separated
from the central $d_{x^{2}-y^{2}}$-wave superconductor by two insulating
thin films. The $d_{x^{2}-y^{2}}$-wave superconductor is presumably
described by the BCS framework. Suppose that the energy relaxation time of
quasiparticles in the superconductor is shorter than the time between two
successive tunneling events, while the latter is shorter than the spin
relaxation time, which ensures that the electrons tunneling into the
superconducting spacer could comply the Fermi distribution, and meanwhile
keep their spin directions. For simplicity, the Andreev reflection effect
will be reasonably ignored, as the resistance of this tunnel junction with
insulating thin films is presumed to be greater than that of a conventional
metallic contact. The schematic layout of the system is shown in Fig. 1.

\begin{figure}[tb]
\begin{center}
\leavevmode\includegraphics[width=0.75\linewidth,clip]{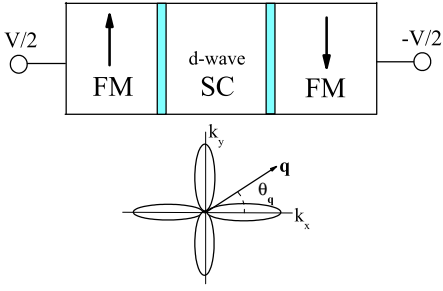}
\end{center}
\par
\vspace{-0.5cm}
\caption{(Color online) The schematic illustration of the F/d-SC/F double
tunnel junction. The lower depiction indicates the $d_{x^{2}-y^{2}}$-wave
gap symmetry in the central superconductor, with $\protect\theta _{\mathbf{q}%
}$ the azimuthal angle of the momentum $\mathbf{q}$.}
\label{fig1}
\end{figure}

In terms of the standard tunneling Hamiltonian and within the linear
response theory, the tunneling current through the $i$th junction can be
obtained by\cite{jin1} 
\begin{equation}
I_{i\sigma }=2\pi e\left\vert \widetilde{T}\right\vert ^{2}D_{i\sigma
}[N-\eta _{i}(\sigma S+\frac{Q-\widetilde{N}}{2})],  \label{current}
\end{equation}%
where$\ \widetilde{T}$\ is the tunneling matrix element,$\ i=1$, $2$, $%
D_{i\sigma }\ $is the subband density of states (DOS) in the $i$th FM
electrode, $\sigma =\pm 1$ for spin up and down, respectively, $\eta _{1}=1$%
, $\eta _{2}=-1$, $S$, $Q$, $N$\ and $\widetilde{N}$\ are given by

\begin{eqnarray}
S &=&\frac{1}{2}\sum_{\mathbf{k}}(f_{\mathbf{k\uparrow }}-f_{\mathbf{%
k\downarrow }}),  \label{spin} \\
Q &=&\sum_{\mathbf{k}}(\text{ }u_{\mathbf{k}}^{2}-v_{\mathbf{k}}^{2}\text{ }%
)(f_{\mathbf{k\uparrow }}+f_{\mathbf{k\downarrow }}),  \label{charge} \\
N &=&\frac{1}{2}\sum_{\mathbf{k}}[f_{0}(E_{\mathbf{k}}-\frac{eV}{2}%
)-f_{0}(E_{\mathbf{k}}+\frac{eV}{2})],  \label{num-1} \\
\widetilde{N} &=&\sum_{\mathbf{k}}(u_{\mathbf{k}}^{2}-v_{\mathbf{k}%
}^{2})[f_{0}(E_{\mathbf{k}}-\frac{eV}{2})+f_{0}(E_{\mathbf{k}}+\frac{eV}{2}%
)].  \label{num-t}
\end{eqnarray}%
In these above equations, $f_{0}(z)$\ is the Fermi distribution function of
thermal equilibrium, and $f_{\mathbf{k}\sigma }$ ($\sigma $ $=\uparrow
,\downarrow $) is the nonequilibrium distribution function of
quasiparticles. The excitation energy is given by $E_{\mathbf{k}}=\sqrt{\xi
_{\mathbf{k}}^{2}+\Delta _{\mathbf{k}}^{2}}$\ for the central $%
d_{x^{2}-y^{2}}$-wave superconductor. The $d_{x^{2}-y^{2}}$-wave SC gap
parameter is defined as usual by $\Delta _{\mathbf{k}}=\Delta \cos (2\theta
_{\mathbf{k}})$, where $\theta _{\mathbf{k}}=\tan ^{-1}(k_{y}/k_{x})$ is the
azimuthal angle of the momentum $\mathbf{k}$, and $\xi _{\mathbf{k}}$ is the
single-particle energy measured from the Fermi level. The coherence factors\ 
$u_{\mathbf{k}}$ and $v_{\mathbf{k}}$ are determined by $u_{\mathbf{k}}^{2}=%
\frac{1}{2}(1+\xi _{\mathbf{k}}/\sqrt{\xi _{\mathbf{k}}^{2}+\Delta _{\mathbf{%
k}}^{2}})$, and $v_{\mathbf{k}}^{2}=\frac{1}{2}(1-\xi _{\mathbf{k}}/\sqrt{%
\xi _{\mathbf{k}}^{2}+\Delta _{\mathbf{k}}^{2}})$. $S$\ and $Q$\ denote the
spin density and the quasiparticle charge density, respectively.

\textit{FFLO state in a }$d_{x^{2}-y^{2}}$\textit{-wave superconductor.}
---As mentioned above, in the case of antiparallel alignment of
magnetizations, the nonequilibrium spin accumulation near the interfaces of
the tunnel junction can appear\cite{takahashi,jin1}, thereby giving rise to
a chemical potential shift that plays essentially the same role as the
spin-exchange field explored by FFLO in their seminal articles\cite{FF,LO}.
Since the tunnel junction under interest is symmetrical, we may presume the
chemical potential shift $\delta \mu _{\uparrow }=-\delta \mu _{\downarrow
}=\delta \mu $\cite{takahashi,jin1}. Generally, the FFLO state involves the
finite momentum $\mathbf{q}$ pairing of electrons\cite{maki}. In the
following, we will consider the FFLO state just in the sense of the Cooper
pairing with nonzero momentum $\mathbf{q}$. For a $d_{x^{2}-y^{2}}$-wave gap
symmetry with finite momentum $\mathbf{q}$ pairing, the energy of
quasiparticles in the presence of the chemical potential shift induced by
the nonequilibrium spin accumulation has the form of $E_{\mathbf{k\sigma }}=$%
\ $\sqrt{\xi _{\mathbf{k}}^{2}+\Delta _{\mathbf{k}}^{2}}+\sigma \lbrack 
\frac{v_{F}\left\vert \mathbf{q}\right\vert }{2}\cos (\theta _{\mathbf{k}}-$ 
$\theta _{\mathbf{q}})-$ $\delta \mu ]$, where $v_{F}$ is the Fermi
velocity. The amplitude $\Delta $\ is determined by the gap equation%
\begin{eqnarray}
1 &=&V_{BCS}N(0)\int_{0}^{\varpi _{c}}d\xi _{\mathbf{k}}\int_{0}^{2\pi }%
\frac{d\theta _{\mathbf{k\ }}}{2\pi }\frac{\cos ^{2}(2\theta _{\mathbf{k}})}{%
2E_{\mathbf{k}}}[\tanh (\frac{E_{\mathbf{k\uparrow }}}{2T})  \notag \\
&&+\tanh (\frac{E_{\mathbf{k\downarrow }}}{2T})],  \label{gap-eq}
\end{eqnarray}%
where $V_{BCS}$\ is the BCS-type pair interaction, $N(0)$ is the DOS in the
normal state at the Fermi level, and $\varpi _{c}$ is the cutoff energy. In
the absence of spin-flip scattering, the spin up and down tunneling currents
should be conserved, i.e. $I_{1\sigma }=I_{2\sigma }$,\ leading to%
\begin{eqnarray}
S &=&PN,  \label{S-A} \\
Q &=&0,  \label{Q-FA}
\end{eqnarray}%
where $P=|D_{i\uparrow }-D_{i\downarrow }|/|D_{i\uparrow }+D_{i\downarrow }|$
is the spin polarization of the FM electrodes. Eq. (\ref{Q-FA}) tells that
there is no charge accumulation in the system. Following the treatments in
Refs.\cite{takahashi,jin1}, we consider the solution of form 
\begin{equation}
f_{\mathbf{k\sigma }}=f_{0}(E_{\mathbf{k}}-\sigma \delta \mu ).  \label{fF}
\end{equation}%
It is applicable when the thickness of the central superconductor is much
smaller than the spin diffusion length, and the spin relaxation time is
sufficiently long\cite{note}. These above equations should be solved
self-consistently.

As discussed by Abrikosov\cite{abrikosov}, for an inhomogeneous
superconducting state, the nonzero solution for $\Delta $\ implies only the
local minimum of the free energy. For the multiple solutions for different $%
\mathbf{q}$, only the value of $\mathbf{q}$ (thus $\theta _{\mathbf{q}}$)
that gives the lowest free energy of the system is kept. The free energy of
the present system bears the following form\cite{heslinga}%
\begin{equation}
F_{S}-F_{N}=\frac{1}{2}\int_{0}^{\Delta }d\Delta ^{\prime }\Delta ^{\prime 2}%
\frac{d}{d\Delta ^{\prime }}\left\{ \sum_{\mathbf{k}}\frac{(\cos 2\theta _{%
\mathbf{k}})^{2}}{E_{\mathbf{k}}}[1-f(E_{\mathbf{k}\uparrow })-f(E_{\mathbf{k%
}\downarrow })]\right\}
\end{equation}%
where $F_{S}$ and $F_{N}$ represent the free energy of the SC state and the
normal state, respectively. By comparing the free energies of different
phases, the phase diagram of the central superconductor can be obtained.

\textit{Phase diagram.} ---Figure 2 presents the phase diagram of the
central $d_{x^{2}-y^{2}}$-wave superconductor in the temperature ($T/T_{c}$%
)-bias voltage ($eV/2\Delta _{0}$) plane for $P=0.4$, where $T_{c}$ is the
SC transition temperature, and $\Delta _{0}=\Delta (T=0)$. Along $V_{c}$
line, the superconductivity is completely suppressed. It turns out that the $%
V_{c}$ line is the phase boundary that separates the SC states from the
normal state, and consequently, the phase transition along the $V_{c}$ line
is of second-order. Below the $V_{c}$ line, there are three phases in four
regions: (i) the d-wave BCS phase, which is in a homogeneous $%
d_{x^{2}-y^{2}} $-wave BCS superconducting state with zero momentum pairing;
(ii) the region encompassed by C$_{1}$B$_{1}$B$_{2}$C$_{1}$, which is in an
inhomogeneous FFLO state with finite momentum pairing at $\theta _{\mathbf{q}%
}=0$; (iii) the region encompassed by C$_{1}$B$_{2}$B$_{3}$C$_{2}$LPC$_{1}$,
which is in the inhomogeneous FFLO state with finite momentum pairing at $%
\theta _{\mathbf{q}}=\pi /4$; (iv) the small region encompassed by C$_{2}$B$%
_{3}$V$_{QCP}$C$_{2}$, which is in the FFLO state at $\theta _{\mathbf{q}}=0$%
, the same phase as in (ii). The phase boundaries, LPC$_{1}$B$_{1}$, C$_{1}$B%
$_{2}$, and C$_{2}$B$_{3}$, separate these four regions, along which the
phase transitions between the homogeneous d-wave BCS state and the
inhomogeneous FFLO states with $\theta _{\mathbf{q}}=0$ and $\pi /4$ are of
the first-order. This statement is clearly confirmed in Fig. 3, where the
chemical potential shift induced by the spin accumulation versus the bias
voltage shows two discontinuities for $\theta _{\mathbf{q}}=0$ and $\pi /4$
at specific voltages. The inset of Fig. 3 presents the gap amplitude $\Delta
/\Delta _{0}$ versus the bias voltage $eV/2\Delta _{0}$, which also shows
that the transitions from the d-wave BCS state to the FFLO state with $%
\theta _{\mathbf{q}}=0$, and then to the FFLO state with $\theta _{\mathbf{q}%
}=\pi /4$ are discontinuous. These discontinuities characterize the
first-order phase transitions between those different SC states. With
increasing the bias voltage, the chemical potential shift is increasing in
trend and coincides with that in normal state, while the gap amplitude is
decreasing and vanishes at $V_{c}$, as shown in Fig. 3. Our calculations
demonstrate that no other stable FFLO states with $\theta _{\mathbf{q}}$\
other than $0$ and $\pi /4$ can be found. In another word, the FFLO phases
with $\theta _{\mathbf{q}}=0$ and $\pi /4$ have the lowest free energies
than those with any other values of $\theta _{\mathbf{q}}$. 
\begin{figure}[tb]
\begin{center}
\leavevmode\includegraphics[width=0.75\linewidth,clip]{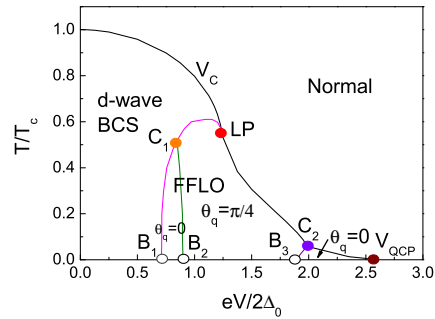}
\end{center}
\par
\vspace{-0.5cm}
\caption{(Color online) The phase diagram of the central $d_{x^{2}-y^{2}}$%
-wave superconductor in the temperature-bias voltage plane for the
antiparallel alignment of magnetizations at $P=0.4$. Four phases are
observed: the homogeneous $d_{x^{2}-y^{2}}$-wave BCS state with zero
momentum pairing, the FFLO states with finite momentum pairing $\mathbf{q}$
for $\protect\theta _{\mathbf{q}}=0$ and $\protect\pi /4$, and the normal
state. A Lifshitz point and a bicritical point are identified, and a quantum
phase transition is sepcified at $V_{QCP}$.}
\label{fig2}
\end{figure}

In the phase diagram, there are three characteristic points along $V_{c}$
line: LP, C$_{2}$ and V$_{QCP}$. At the point LP, a disordered phase (normal
state), an ordered phase (d-wave BCS state) and a spatially modulated phase
(FFLO state with $\theta _{\mathbf{q}}=\pi /4$) meet, showing that the point
LP is a Lifshitz point\cite{lifshitz}. Although the Lifshitz point was
originally proposed for a metamagnet, we now have another example in a
different system. At the point C$_{2}$, two second-order phase transition
lines meet with a first-order phase transition line, suggesting that C$_{2}$
is a bicritical point. At the point V$_{QCP}$, satisfying $PeV_{QCP}/2\Delta
_{0}=1.03$, a quantum phase transition (QPT) from the ordered FFLO state
with $\theta _{\mathbf{q}}=0$ to the disordered normal state happens at $T=0$%
, indicating that V$_{QCP}$ is a quantum critical point. Such a QPT is of
second-order.

At the point C$_{1}$, three first-order phase transition lines meet,
implying that the three SC phases can coexist. Thus, C$_{1}$ is a triple
transition point. At the points B$_{1}$, B$_{2}$, B$_{3}$, the first-order
phase transitions may appear at $T=0$, namely, at B$_{1}$ the homogeneous
d-wave BCS state coexists with the inhomogeneous FFLO state with $\theta _{%
\mathbf{q}}=0$; and at B$_{2}$, B$_{3}$, the FFLO states with $\theta _{%
\mathbf{q}}=0$ and $\theta _{\mathbf{q}}=\pi /4$ can coexist in the ground
state. These first-order phase transitions at zero temperature are driven by
external voltages, not by thermal fluctuations. 
\begin{figure}[tb]
\begin{center}
\leavevmode\includegraphics[width=0.75\linewidth,clip]{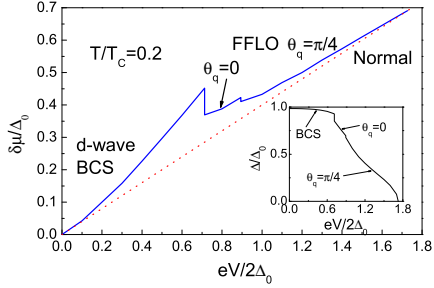}
\end{center}
\par
\vspace{-0.5cm}
\caption{(Color online) The bias dependence of the chemical potential shift
induced by the spin accumulation at $P=0.4$, $T/T_{c}=0.2$, where two
discontinuities are observed. Inset: the bias voltage dependence of the gap
amplitude ($\Delta /\Delta _{0}$).}
\label{fig3}
\end{figure}

\textit{Tunnel conductance and magnetoresistance.} ---The tunnel currents
are given by $I^{A}=I_{0}(1-P^{2})N$ for the antiparallel (A) alignment of
magnetizations, and $I^{P}=I_{0}N$ for the parallel (P) alignment of
magnetizations, where $I_{0}=2\pi e\left\vert \widetilde{T}\right\vert
^{2}(D_{i\uparrow }+D_{i\downarrow })$. The tunnel conductance can be
obtained by $G^{A,P}=dI^{A,P}/dV$, and the tunnel magnetoresistance (TMR)\
is defined by $TMR=\frac{G^{P}}{G^{A}}-1$. Fig. 4 gives the tunnel
conductance $G_{S}^{A,P}/G_{N}^{A}$ versus the bias voltage ($eV/2\Delta
_{0} $) at $T/T_{c}=0.2$ for $P=0.4$, where $G_{N}^{A}$ is the tunnel
conductance in the normal state, and the curves of the conductance
corresponding to different SC phases are indicated. It can be seen that the
conductance $G_{S}^{A}$ in the A configuration shows exotic behaviors with
jumps and kinks, in comparison to the conductance $G_{S}^{P}$ in the P
configuration, which reflects the features of the superconducting density of
states. This behavior is also different from the s-wave superconductor
explored in Ref.\cite{jin1}. Fig. 5 shows the bias voltage dependence of the
tunnel TMR. One may observe that the TMR displays complex behaviors with
cusps and kinks, being resulted from the first-order phase transitions
between different SC phases. The behavior of TMR is also obviously different
from the s-wave superconductor discussed in Ref.\cite{jin1}. The results of
the tunnel conductance and TMR show that the anisotropy of the SC gap
parameter indeed has an essential effect on the spin-dependent transport of
electrons in F/SC/F double tunnel junctions. 
\begin{figure}[tb]
\begin{center}
\leavevmode\includegraphics[width=0.75\linewidth,clip]{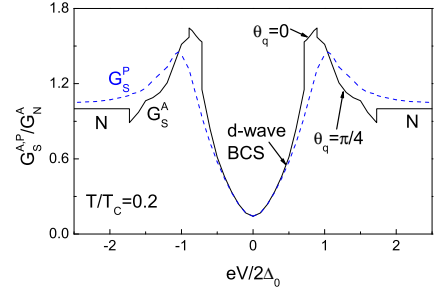}
\end{center}
\par
\vspace{-0.5cm}
\caption{(Color online) The bias voltage dependence of the tunnel
conductance at $P=0.4$, $T/T_{c}=0.2$, where N denotes the normal state.}
\label{fig4}
\end{figure}
\begin{figure}[tb]
\begin{center}
\leavevmode\includegraphics[width=0.75\linewidth,clip]{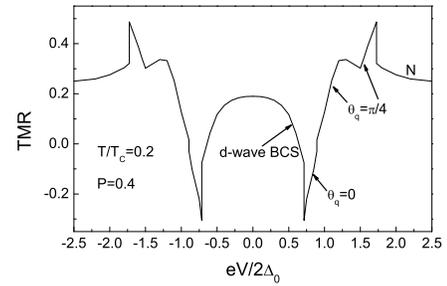}
\end{center}
\par
\vspace{-0.5cm}
\caption{The bias voltage dependence of the tunnel magnetoresistance at $%
P=0.4$, $T/T_{c}=0.2$, where N denotes the normal state.}
\label{fig5}
\end{figure}

\textit{Summary. ---}In summary, we have theoretically investigated the
effect of spin accumulation on the spin-dependent transport in F/d-SC/F
double tunnel junctions. The FFLO inhomogeneous SC state, first- and
second-order phase transitions, and quantum criticality induced by the spin
accumulation in the central $d_{x^{2}-y^{2}}$-wave superconductor are
predicted. A complex phase diagram in the temperature-bias voltage plane is
determined. It is shown that the phase transitions from the homogeneous BCS
state with zero momentum pairing to the inhomogeneous FFLO state with finite
momentum pairing, and from the FFLO state with $\theta _{\mathbf{q}}=0$ to
that with $\theta _{\mathbf{q}}=\pi /4$, are of the first-order; while the
transitions from all SC states to the normal state at critical voltages $%
V_{c}$ are of the second-order. A Lifshitz point, a bicritical point and a
quantum critical point are identified. The tunnel conductance and TMR are
also obtained. It is found that the effect of the spin accumulation on the
spin-dependent transport in F/d-SC/F double tunnel junctions is quite
different from that in F/s-SC/F tunnel junctions.

\acknowledgments

This work is supported in part by the NSFC (Grant Nos. 90403036, 20490210,
10247002).


\begin{references}
\bibitem{note}Note that a long spin-diffusion length is a necessary but not a sufficient condition for the form of Eq. (9).
\end{references}
\end{document}